\documentclass[10pt]{spie}
\usepackage[utf8]{inputenc}
\usepackage{authblk}
\usepackage[square,sort,comma,numbers]{natbib}
\usepackage{graphicx}
\usepackage{float}
\usepackage{gensymb}
\usepackage{url}
\usepackage{caption}
\usepackage{subcaption}
\usepackage{amssymb}
\usepackage{amsmath}
\usepackage{bm}
\usepackage{siunitx}
\usepackage[normalem]{ulem}
\usepackage{wrapfig}
\usepackage{color}
\usepackage{upgreek}
\usepackage{booktabs}
\usepackage{multirow}
\usepackage{pbox}

\usepackage{geometry}
\title{Slit Mask Integral Field Units for the Southern African Large Telescope}
\author[a,*]{Sabyasachi Chattopadhyay}
\author[a,b,c]{Matthew A. Bershady}
\author[b]{Marsha J. Wolf}
\author[b]{Michael P. Smith}

\affil[a]{South African Astronomical Observatory, 1 Observatory Rd, Observatory, Cape Town, 7925, South Africa}

\affil[b]{University of Wisconsin, Department of Astronomy, 475 North Charter Street, Madison, WI 53706, USA}

\affil[c]{Department of Astronomy, University of Cape Town, Private Bag X3, Rondebosch 7701, South Africa}

\begin{document}

\maketitle

\begin{abstract}

Two fibre integral field units (IFU) are being built in the SAAO fibre-lab for the Robert Stobie Spectrograph's visible arm and the future red arm.  Each IFU sits in its own slit-mask cassette and is referred to as a slit-mask IFU (SMI). They will be available some time in 2022. The smaller, 200 micron fibre IFU has 309 x 0.9 arcsec diameter spatial elements covering an elongated hexagonal footprint of 18 x 23 arcsec. The larger, 400 micron fibre IFU has 178 x 1.8 arcsec diameter spatial elements covering an on-sky area of 21 x 44 arcsec. In both cases there are two groups of 13 fibres offset by roughly 50 arcsec on either side of the primary array to sample sky. The 1.8 and 0.9 arcsec spatial resolution SMIs provide median spectral resolution of 1200 and 2400 respectively at H-$\alpha$ wavelengths in the low resolution mode covering 320 to 740 nm bandpass. At a higher grating angle the SMI will deliver spectral resolution up to 5000 and 10000 with 400 and 200 micron core fibre respectively. A future red-arm will extend the simultaneous wavelength coverage up to 900 nm at a median resolution of 3000/6000 for the same flavors of IFUs. SMIs are inserted in the same fashion as the existing long-slit cassettes at the SALT focal plane. Prismatic fold mirrors direct the focal plane into the fibre IFU and then back into the RSS collimator after the fibres are routed 180 deg within the cassette and formatted into a pseudo-slit. Fold-prisms ensure that the spectrograph collimator continues to see the same focal plane. In this paper we describe the design, fabrication, assembly and characterization of Slit Mask IFUs.
\end{abstract}


\section{Introduction}
Fibre-fed integral field spectroscopy (IFS) has evolved from its nascent stages \cite{Allington-Smith_2002,bershady,giraffe} and is now a main-stream observational mode on telescopes around the world \cite{kelz,virusp,virusw,SAMI,Sanchez,manga,vlt,kmos,muse,vimos}. To date, SALT does not have such capability; work presented here addresses that shortfall. Given the photon capturing power of large telescopes, IFS can be key for observing extended fainter diffused sources such as ISM, nebulae, globular clusters in and around a galaxy \cite{cwi,manga,Sanchez,SAMI}. The power of fibre fed IFS lies in the ability in reformatting a 2D sky area into 1D pseudo slit. Except for scientific cases where high stability of spectra is necessary, integral field units (IFU) can be co-mounted with the spectrograph on the telescope prime focus or Cassegrain port. However, modifying these ports to include a new instrument along with the existing long slits can be challenging and costly. Often observers wants to switch between IFU and long slit observation for the choices of slit width and throughput despite the limitations on the later. Hence an IFU of the similar scale to that of an long slit can be very useful to be used interchangeably with the long slits. This scheme ensures instrument development costs are small (in contrast to full-fledged spectrograph optics and detectors), and that instrument installation minimizes or eliminates telescope down-time.

The South African Large Telescope (SALT) already hosts the Robert Stobie Spectrograph (RSS) with a suite of long-slit and multi-slit masks for its focal plane. We are developing multiple Slit Mask IFU (SMI) for the SALT focal plane following the principle of augmenting existing spectrographs, described above. In this paper we begin by presenting the the scientific capabilities of these instrument on SALT. We then describe the SMI instrument components and the the critical process development required for routing and polishing the fibres. This is followed by a presentation of fibre performance measurements and methodology. We conclude with summary of results along with the current status.

\section{Scientific capabilities}

\begin{figure}[H]
    \centering
    \includegraphics[width=0.8\linewidth]{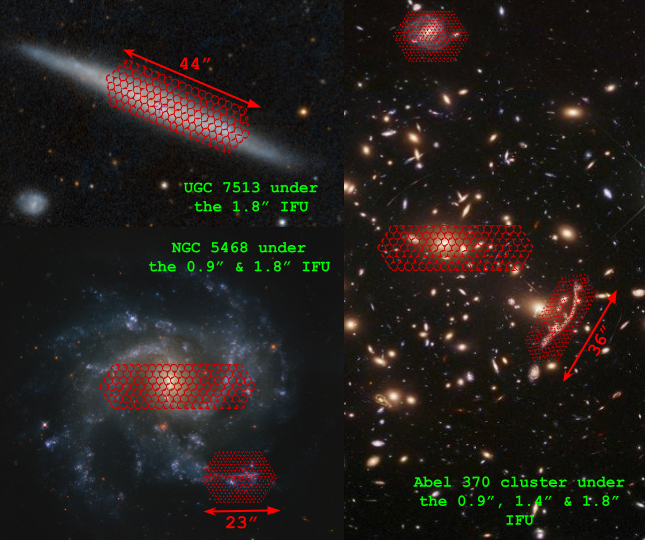}
    \caption{Footprint of different Slit Mask IFUs on sky.}
    \label{fig:IFUFootprint}
\end{figure}

The elongated hexagonal shapes of SMI, shown in figure \ref{fig:IFUFootprint}, are ideal for observing galaxies over a range of inclination angles, and can be used to map more extended objects from Galactic HII regions, to merging and interacting galaxies, to galaxy cluster cores and strongly lensed galaxies. Additionally, the SMIs can be effective as a photon bucket for low surface brightness observations as well.

\begin{figure}[H]
    \centering
    \includegraphics[width=0.8\linewidth]{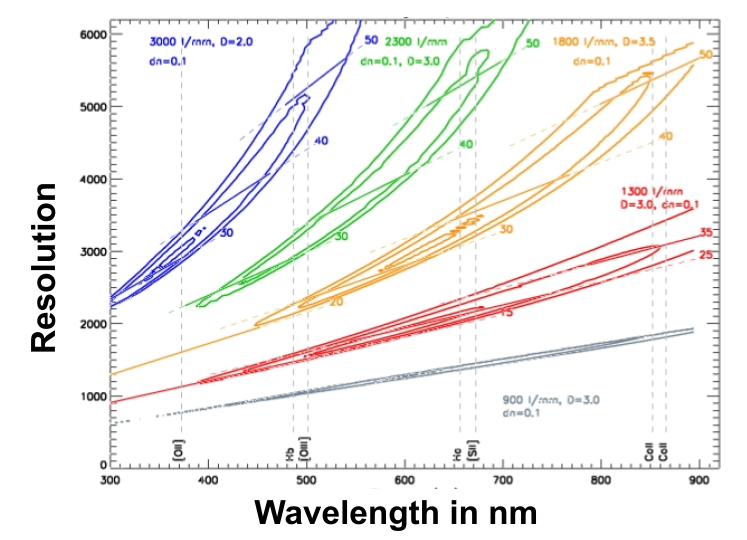}
    \caption{Spectral resolution of Robert Stobie Spectrograph with different grating configurations for a 1.5 arcsec slit-width, close to the effective width for the large-fibre IFU.}
    \label{fig:Resolution}
\end{figure}

The SMIs will sit in front of the Robert Stobie Spectrograph (RSS). RSS has an 8 arcmin field and wavelength coverage of 320-900nm. The wavelength range will be dividable into blue and red channels of 320nm-650nm and 650nm-900nm, respectively, with the addition of red arm in coming years. RSS employs five volume phase holographic gratings (VPHg) that are rotated to achieve the full wavelength coverage. The VPHg's have 700, 1300, 1800, 2300 and 3000 lines/mm; when operated at 12$\degree$, 19$\degree$, 30$\degree$, 35$\degree$ and 50$\degree$, respectively, their central wavelength is 500nm. A detailed view of spectral resolution and wavelength is provided in Figure~\ref{fig:Resolution}. RSS has a mini-mosaic of three abuttable E2V 44-82 CCDs with 2k$\times$4k 15 $\mu$m pixels. RSS produces a total of 914 spectral elements for 0.9 arcsec wide slit. Inclusion of the red arm will double the number of resolution elements simultaneously sampled.

The 200um fibre IFU contains 309 spatial elements covering an extended hexagonal footprint of 17.86$\times$22.92 arcsec on the sky. Each fibre collects light from a 0.9 arcsec diameter circular aperture. The hexagonal fibre array has a fill-factor of roughly 57.5\%. This fill-factor arises from the hexagonal packing and the 240 um (1.06 arcsec) centre-to-centre spacing required by the fibre clad and buffer. To achieve 100\% spatial coverage, a series of three observations can be taken with the array dithered on sky in three defined positions. There are an additional 27 fibres in two offset locations to sample sky on either side of the hexagonal array. The sky fibres are arranged in two linear arrays, each spanning 15 arcsec, oriented on the long axis of the central hexagonal array. Each linear array is offset from the hexagonal array by roughly 70 arcsec (center to center), with edge-to-edge spacing of at least 50 arcsec. For objects that fill both the central hexagonal array and the sky arrays, it is necessary to have separate, offset exposures to sample sky. The 336 fibres are arranged in a pseudo-slit spanning roughly the 8 arcmin ($\sim$108mm) diameter entrance aperture of the spectrograph, each producing a well-separated spectrum that will be extracted and wavelength calibrated as separate spectra. To optimize sky-subtraction, the sky fibres are carefully interleaved with the object fibres along the slit, and are position on sky to mimic the telecentric angle of the object fibres. The spectral resolution will be equivalent to an 0.78 arcsec width long-slit given the circular entrance aperture geometry.

The larger 400 um fibre IFU contains 178 spatial elements covering an extended hexagonal footprint of 20.78$\times$41.43 arcsec. Each fibre collects light from a 1.8 arcsec diameter circular aperture. The hexagonal fibre array has a similar fill-factor of 60.8\%. The centre-to-centre fibre spacing is 480 um (2.1 arcsec). There are an additional 26 fibres in two offset locations to sample sky, arranged and spaced from the central array similarly to the smaller IFU. The 204 fibres are arranged in a pseudo-slit also in a similar fashion as the smaller-fibre IFU. Spectral resolution and dithering considerations for truly integral spatial coverage are the same as for the smaller IFU, adjusted for the different fibre size and spacing. For a given grating configuration the spectral resolution is reduced by a factor of two. A third, planned IFU will have 300um fibre and footprint identical to the NIR spectrograph IFU soon to be commissioned.

Given the short length of fibre, transmission losses within the fibres are negligible even in the far-blue. While the prisms have been AR-coated, the fibres are not, so there are two air-glass surface-losses of roughly 3.4\% each. In addition, the fibres degrade in the input beam profile (focal ratio degradation - FRD), such that the RSS collimator becomes slightly over-filled. The combination of these effects is estimated to amount to a 15\% throughput loss. However, the scrambling properties of the fibres should minimize illumination variations over a SALT track, improving sky-subtraction performance. 

\section{Instrument Description}

\begin{figure}[H]
    \centering
    \includegraphics[width=\linewidth]{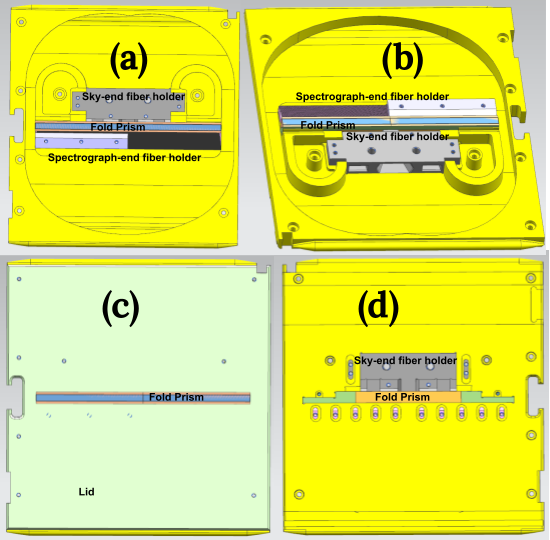}
    \caption{Model of the Slit Mask IFU cartridge external and internal structure without fibre: (a) Top view without lid; (b) Tilted top view; (c) Top view with lid; (d) Bottom view.}
    \label{fig:SlitMaskIFU}
\end{figure}

SALT prime focus module currently hosts a series of long slit and multi aperture masks with different width to match the on sky seeing. These masks sits in a cartridge case from which a robotic arm picks them and insert at the focal plane. SMIs would join these range of masks in the cartridge case as a feature instrument. Consequently the outer dimension of the IFU would need to match to that of existing slit masks which is 134mm $\times$ 130mm $\times$ 8mm which is similar to a DVD cassette. This forces the fibres to be laid at the SALT focal plane as shown in Figure \ref{fig:SlitMaskIFU}. The instrument consists of mainly three parts; (i) 2D fibre array at the sky end, (ii) 1D fibre array at the spectrograph end, and (iii) reflecting prisms to fold the optical axis into (at the sky end) and out of (at the spectrograph end) focal plane.

\subsection{Sky-end Aperture}

\begin{figure}[H]
    \centering
    \includegraphics[width=\linewidth]{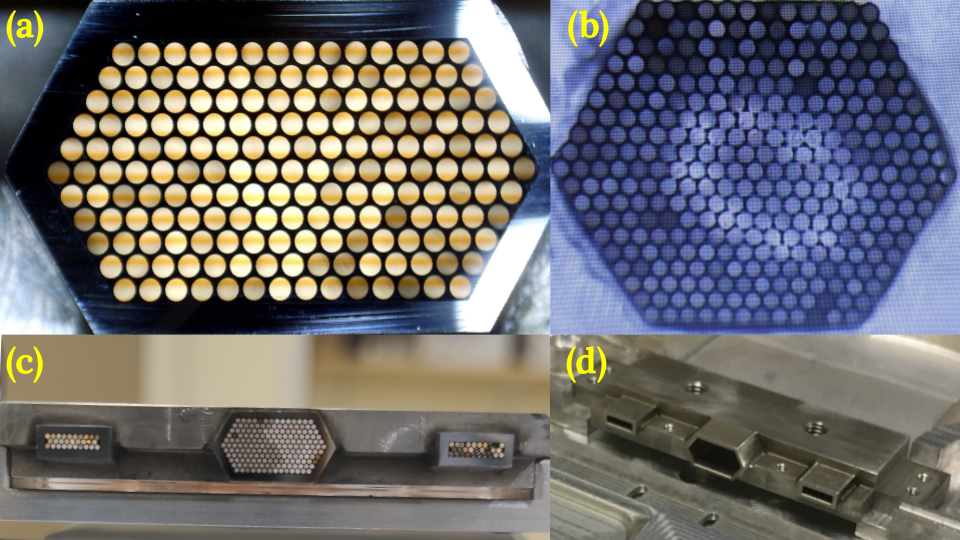}
    \caption{(a) Image of the science version of the fully polished 400 $\mu$m diameter core fibre IFU face. (b) Image of the prototype version of the hand polished 200 $\mu$m diameter core fibre IFU face. (c) Image of object and sky bundle relative position. (d) Empty sky-end fibre holder placed inside the slit mask cartridge.}
    \label{fig:IFUface}
\end{figure}

Fibres are distributed into object and sky bundles. Simultaneous object and sky sampling are particularly critical for a variable aperture telescope such as  SALT. Although it is possible to nod to sky, this would need to be done on a high cadence to sample both temporal variations in the sky and temporal variations of the telescope pupil illumination over a track. Hence the idea of separate sky fibres is adopted. Our design has two bundles of sky fibres on each side of the object bundle. The object 2D array (the `IFU') is shaped in an extended hexagonal aperture with fibres packed in tight honeycomb pattern. The packing is same for the sky bundles but the array is rectangular with three fibre rows. The large format IFU has a total of 204 fibres of 400:440:480um cor:clad:buffer respectively. For the 0.9 arcsec IFU, there are 336 fibres of 200:220:239um. In both cases, object and two sky bundles (on either side of the object bundle) are separated by 80.6 arcsec on sky (center to center between the object and each sky bundle) or 18mm in linear dimensions and their orientation is shown in figure \ref{fig:IFUface}. The 400um (large) IFU has 179 object fibres and 25 sky fibres divided into 12 and 13 for each sky bundle. The 200um (small) IFU has 309 object fibres and 27 sky fibres divided into groups of 13 and 14 similarly. Figure \ref{fig:IFUface} shows the science version of polished large format IFU sky-end and prototype version of the small format IFU.

\begin{figure}[H]
    \centering
    \includegraphics[width=\linewidth]{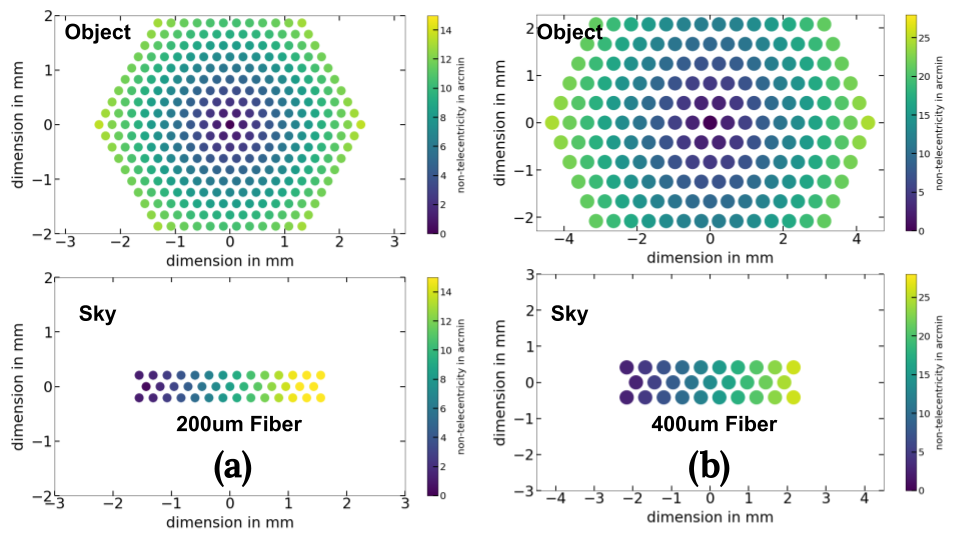}
    \caption{Spatial distribution of non-telecentricity in object and sky fibres for (a) 200 $\mu$m and (b) 400 $\mu$m diameter core fibre IFU. Non-telecentricity is given by the fibre color and bar.}
    \label{fig:SpatialNTDistribution}
\end{figure}

\begin{figure}[H]
    \centering
    \includegraphics[width=0.8\linewidth]{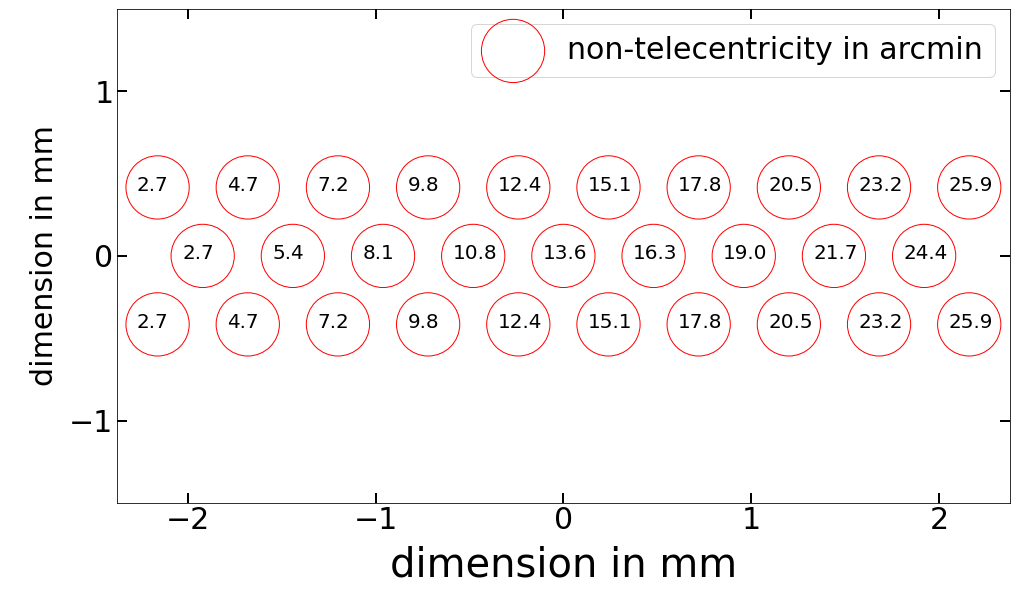}
    \caption{Location and non-telecentricity distribution of full length sky fibres for 400 $\mu$m diameter core fibre IFU. The sky fibres used for slit are marked in bold red.}
    \label{fig:Skyfibres}
\end{figure}

The telescope focal plane has significant  non-telecentricity: 76 arcsec of beam tilt per arcsec offset from the field center on sky. Figure \ref{fig:SpatialNTDistribution} shows the distribution in non-telecentricity of the fibres on both IFUs and their corresponding sky arrays.  The object bundles (IFUs) are at the center of the field, and hence the minimum non-telecentricity is zero with symmetrically increasing in all direction. This yields a modest change in geometric FRD as a function of field-angle from the center. To ensure good sky subtraction it is important that the sky fibres have a comparable range of non-telecentric illuminations (we describe the detailed mapping scheme elsewhere). Hence the sky fibres are tilted at an angle of 1.51$\degree$ and 1.66$\degree$ for large  and small format IFUs respectively. This ensures the \textit{range} of non-telecentricity is similar to that of object fibre bundle as can be seen in figure \ref{fig:SpatialNTDistribution}. To optimize the distribution  of non-telecentricity between object and sky fibres, we only used sky fibres as shown in figure \ref{fig:Skyfibres}; the remainder were of short length for mechanical packing.

As discussed earlier, with RSS, there are five different choices of gratings available which are the following:
\begin{enumerate}
    \item \textbf{900 lines per mm} provides median spectral resolution of 1300/650 at 12 degree from 320nm to 650nm for 1.8/0.9 arcsec fibre. At 20 degree, the achievable wavelength range is 600-900nm at a median resolution of 2200/1100 for same set of fibres. 
    \item \textbf{1300 lines per mm} provides median spectral resolution of 1600/800 at 15 degree from 300nm to 500nm for 1.8/0.9 arcsec fibre. At 31 degree, the achievable wavelength range is 700-900nm at a median resolution of 3600/1900 for same set of fibres.
    \item \textbf{1800 lines per mm} provides median spectral resolution of 3000/1550 at 26 degree from 425nm to 570nm for 1.8/0.9 arcsec fibre. At 50 degree, the achievable wavelength range is 800-900nm at a median resolution of 7200/3700 for same set of fibres.
    \item \textbf{2300 lines per mm} provides median spectral resolution of 3100/1600 at 27 degree from 340nm to 450nm for 1.8/0.9 arcsec fibre. At 48 degree, the achievable wavelength range is 600-700nm at a median resolution of 6800/3500 for same set of fibres.
    \item \textbf{3000 lines per mm} provides median spectral resolution of 3700/1900 at 31 degree from 300nm to 400nm for 1.8/0.9 arcsec fibre. At 48 degree, the achievable wavelength range is 460-530nm at a median resolution of 6700/3500 for same set of fibres.
\end{enumerate}

\subsection{Spectrograph-end pseudo-slit}

\begin{figure}[H]
    \centering
    \includegraphics[width=\linewidth]{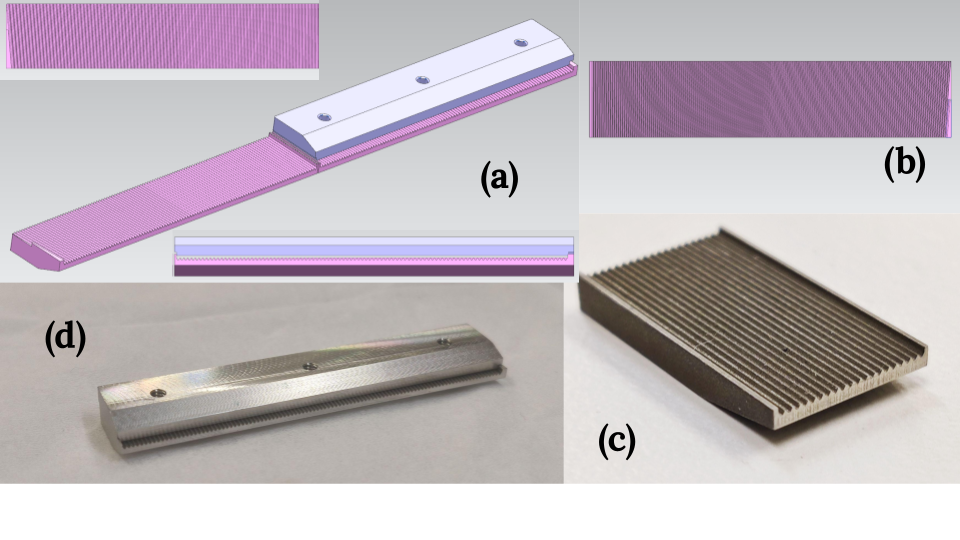}
    \caption{Model of V-groove layout for (a) 400 $\mu$m and (b) 200 $\mu$m diameter core fibre; (c) Test V-groove for 200 $\mu$m IFU; (d) Science V-grooves for 400 $\mu$m IFU.}
    \label{fig:V-groove}
\end{figure}

The visible arm of the RSS (RSS-VIS) at the SALT prime focus accepts a slit length of 480 arcsec or 105.6mm. This length drives the number of fibres i.e. 336 for small format and 204 for large format IFU. The RSS collimator is designed to accept non-telecentric input beam and hence the fibres on the pseudo-slit are tilted away from center to mimic this variation in non-telecentricity. Wire-EDM V-grooves of width 10mm (along the length of the fibre) were fabricated in the SAAO shop with appropriate fan angles to contain the fibres and ensure the continuous variation in non-telecentricity from fibre to fibre. Figure \ref{fig:V-groove} shows the variation in V-groove channel angle in the blocks. The slit is divided into two sub-blocks for the ease of polishing and alignment. The fibres are equally distributed into the two sub-slits. There is a gap of 0.5mm (2.25 arcsec) between the two sub-slits for measurement of stray light. We modelled the effect of fibre separation in the pseudo-slit on cross-talk between adjacent spectra. Given the field-dependent aberrations in the spectrograph, we decided to have different separations between central and edge fibres of the slit. For large format IFU, the inner 50\% if the fibres are separated by 2.22 arcsec (0.494mm) while the outer 50\% are positioned at distance of 2.32arcsec (0.515mm). For the small format IFU, these values are 1.13 arcsec (0.3mm) and 1.19 arcsec (0.32mm) respectively. This separation helps achieving less than 10\% cross talk between adjacent spectra.

\subsection{Fold prisms}

\begin{figure}[H]
    \centering
    \includegraphics[width=0.6\linewidth]{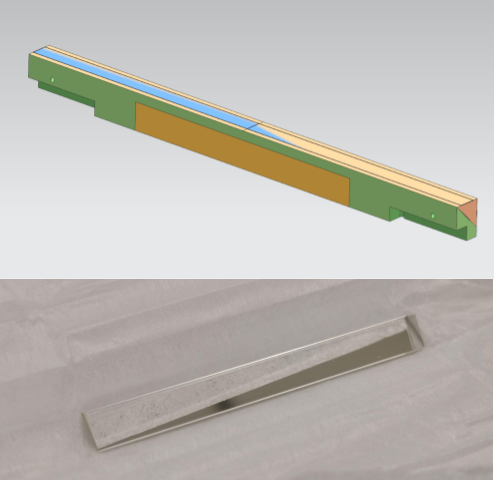}
    \caption{(a) Model of the sky and spectrograph end prism assembly and (b) image of a sky-end prism.}
    \label{fig:FoldPrism}
\end{figure}

The fold prisms is a key component of SMI. These prisms first fold the incoming focal plane from the telescope into the 2D fibre array and then fold the pseudo-slit back into the spectrograph as the outgoing focal plane, or spectrograph entrance aperture. There are three prisms for each SMI; one for folding the input into the object and sky bundles, and two for folding the pseudo-slit into the spectrograph. The prisms also need to be mounted within the 8mm thickness of the IFU. The prisms are arranged to ensure that the location of the focal plane viewed by the spectrograph is unchanged. This helps in keeping the spectrograph setup (i.e., focus) largely unchanged when swapping between modes. The prisms have a mirrored isosceles right triangular cross section (5.8mm leg) along the longer dimension (52.8mm) with AR-coated legs used for input and output of light. The AR coatings have better than 99\% throughput at each surface while the reflective coated hypotenuse used for folding the wavefront has  less than 5\% loss. The vendor, Rocky Mountain Instruments, provided $\lambda$/4 surface finish at 630nm with a clear aperture of 5.8mm$\times$48mm at the sky end and 2mm$\times$52.8mm at the spectrograph end. These dimensions are provided in a graphic form in figure \ref{fig:FoldPrism}.

\section{Fibre Routing and Polishing}

The most critical part of such a compact fibre instrument design is routing of fibres in such a way that they are well protected, fit withing the instrument envelop, and yet are positionable for polishing the fibre surfaces at both the sky (IFU and sky bundle) and spectrograph (pseudo-slit) ends. We will describe how these issues are tackled for SMI below. The basic concept is to assemble the fibres in a fold-able cassette, and then transfer the final assembly to the fixed cassette that inserts into the relevant telescope magazine. The fold-able cassette is essential to access both ends of the fibres for termination and polishing after the precise routing of fibres has been accomplished.

\subsection{Fibre Routing}
\label{subsec:FibreRouting}

Apart from the obvious dimensional restrictions of the cassette, the important requirements of fibre routing are: (i) maintain maximum attainable slit length, (ii) align the pseudo slit and the field center, and (iii) maximize the fibre bend radius. The fibres coming out of the IFU and sky arrays are divided into eight object and 2 sky bundles. Each group is sheathed in thin-wall heat-shrink teflon tubing to avoid bare-fibre cross-over and breakage. These bundles are then  bent 180$\degree$ around one half of the cassette with diameters ranging between 50-63mm. The effective bend radius is very tight, and quite extreme for the 400um fibre in particular.

The fibres then pass through two rectangular channels on either side of the pseudo-slit at the mid-point of the cassette. The channels have cross sectional dimensions of 5.69mm$\times$6.25mm on one side, and 7.3mm$\times$6.25mm on the other. Requirement (ii), the available volume within the slit-mask cassette, and the need for structural robustness drives the two side channels to be slightly different.
% The sub-bundles are covered with teflon tubes outside and inside the side channel to protect the fragile fibres from damage due to friction against the sharp corners and the side walls of the side channels. 

\begin{figure}[H]
    \centering
    \includegraphics[width=0.7\linewidth]{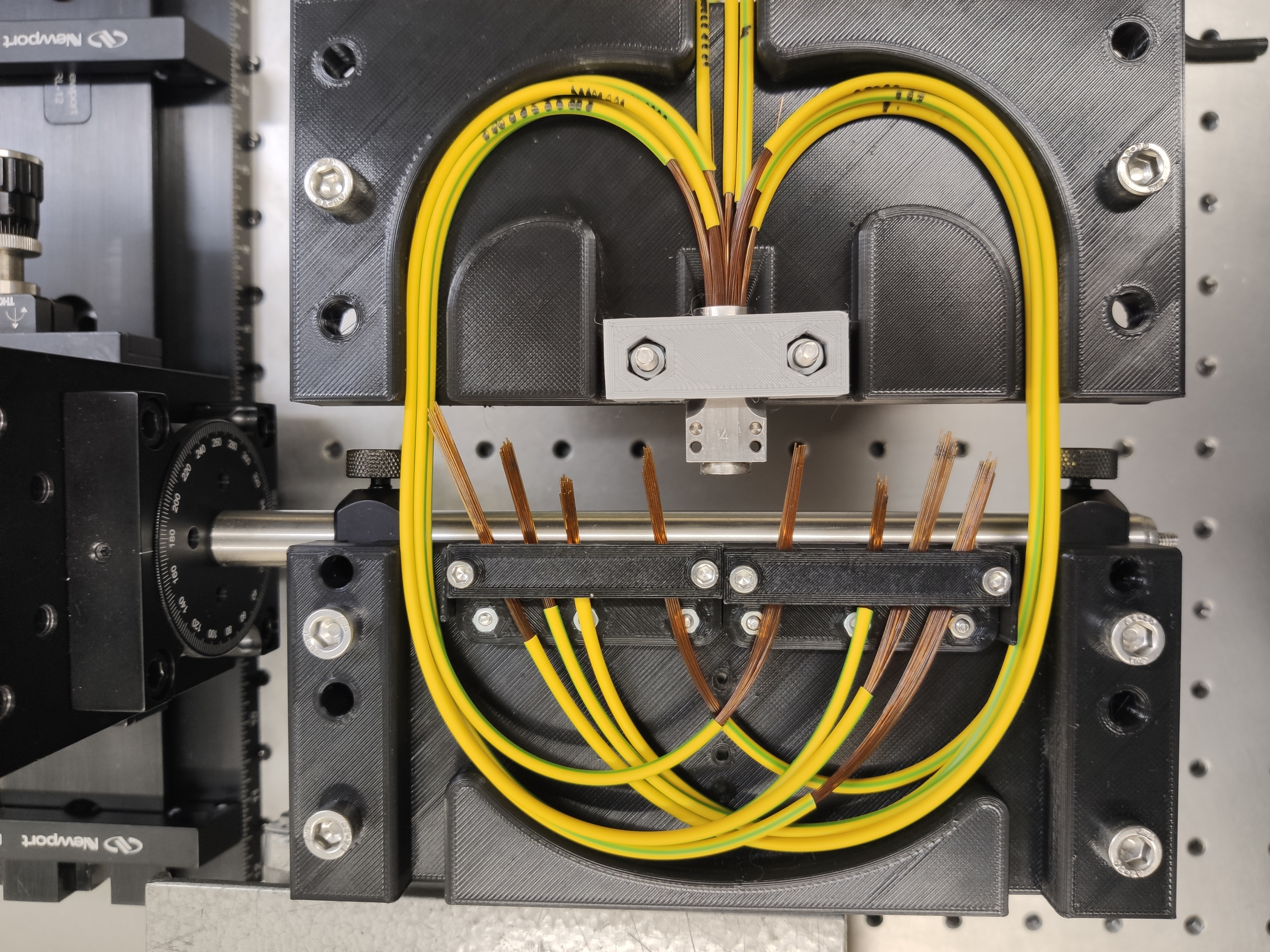}
    \caption{Routing of object and sky fibres inside the 3D printed assembly. The figure depicts the process development version performed with 300 $\mu$m diameter core fibre.}
    \label{fig:FibreRouting}
\end{figure}

Finally, after emerging from the narrow channels, the bundles open up into the slit-half of the cassette. Here they take another 180$\degree$ turn with a radius of 85mm to enter the V-grooves. To achieve this configuration, the sub-bundles are first laid straight pointing away from the V-groove mount location. The fibres are inserted inside the V-groove and the fibre filled V-grooves are slowly turned into the slit location and fastened. The process development version of this configuration is shown in figure \ref{fig:FibreRouting}.

\subsection{Fibre Polishing}

\begin{figure}[H]
    \centering
    \includegraphics[width=0.7\linewidth]{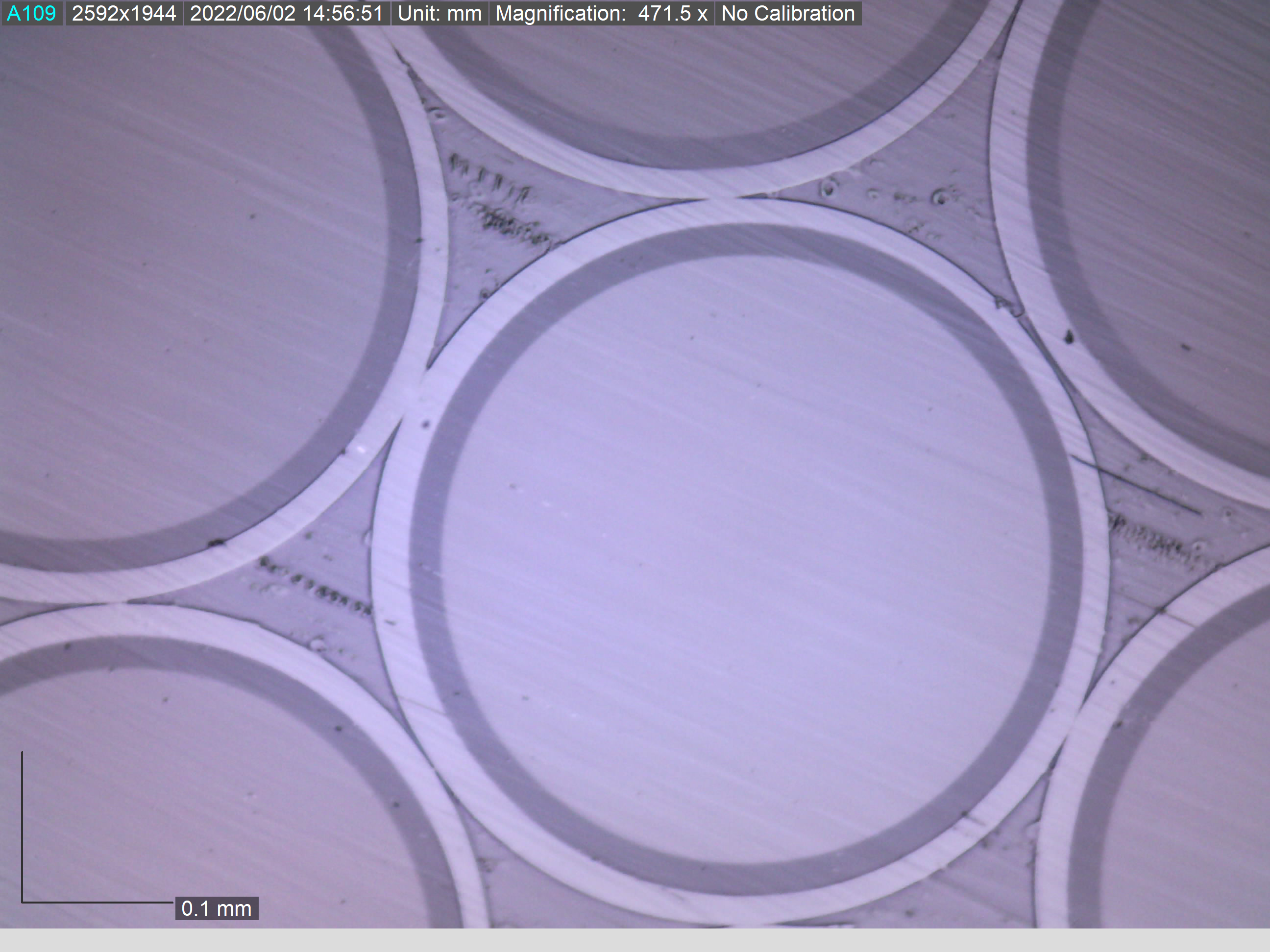}
    \caption{Typical end-finish after polishing object fibres in 400 $\mu$m diameter core fibre IFU.}
    \label{fig:FibreEndFinish}
\end{figure}

The object and sky bundles form a single block and are polished simultaneously using the recipe described in \cite{polisher}. First a length of 600mm long fibres are inserted inside the fibre holder and then two-part epoxy glue (Epotek 301) is used to bond the fibres into the holder. The glue was pre-cured for 2.5 hours to get a slightly higher viscosity to ensure less than 5mm wick back along the length of the packed fibres. The short wick back-length helps in keeping the end stress to be minimal which is known to cause focal ratio degradation \cite{carasco,arthur}. The fibre bundles are then polished using a custom polishing arm on a flat rotating polisher down to 1 micron surface finish. Polished object bundle fibres under axial illumination are shown in figure \ref{fig:FibreEndFinish}. After polishing is complete, the fibres are routed as described in subsection \ref{subsec:FibreRouting}. 

\begin{figure}[H]
    \centering
    \includegraphics[width=0.7\linewidth]{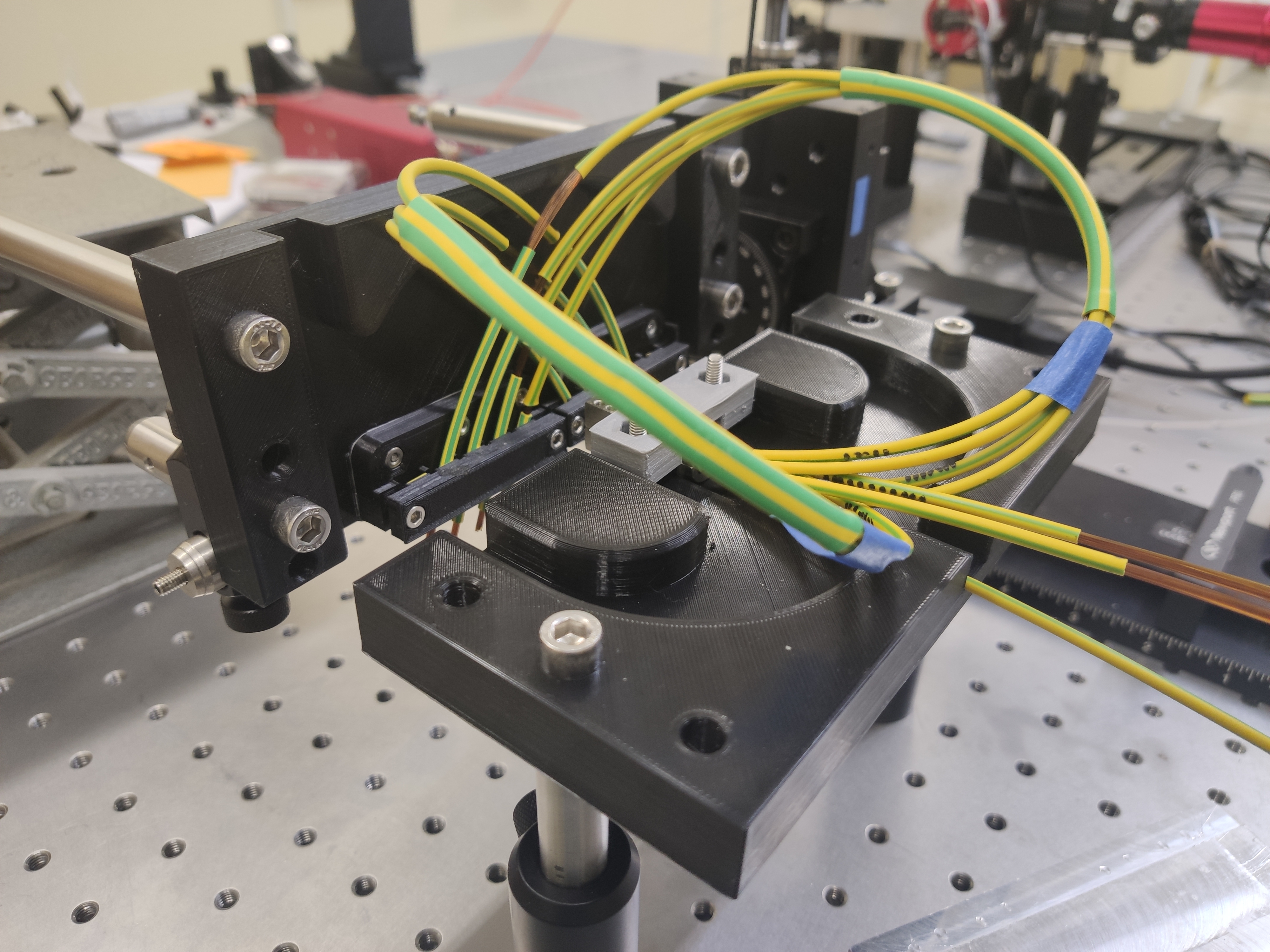}
    \caption{(a) Routing of object and sky fibres inside the 3D printed polishing jig. The figure depicts the process development version performed with 300 $\mu$m diameter core fibre.}
    \label{fig:FibreFolding}
\end{figure}

Once the slit V-groove blocks are fastened in their place, the fibres are glued with similar protocols and epoxy as discussed above. The custom routing and polishing jig is made of two half plates that can be folded up to 90$\degree$, as shown in Figure~\ref{fig:FibreFolding}. The custom jig with glued V-groove is brought to a folded configuration to clear the slit up for polishing. Similar polishing method to that of sky-end is used to polish the slit. Polished slit fibres under circumferential and axial illumination are going to be inspected for end finish as shown in figure \ref{fig:FibreEndFinish}. At the end of polishing, the assembly is then transferred from custom polishing jig to final mask via a transfer plate.

\section{Fibre performance characterization}

\begin{figure}[H]
    \centering
    \includegraphics[width=0.95\linewidth]{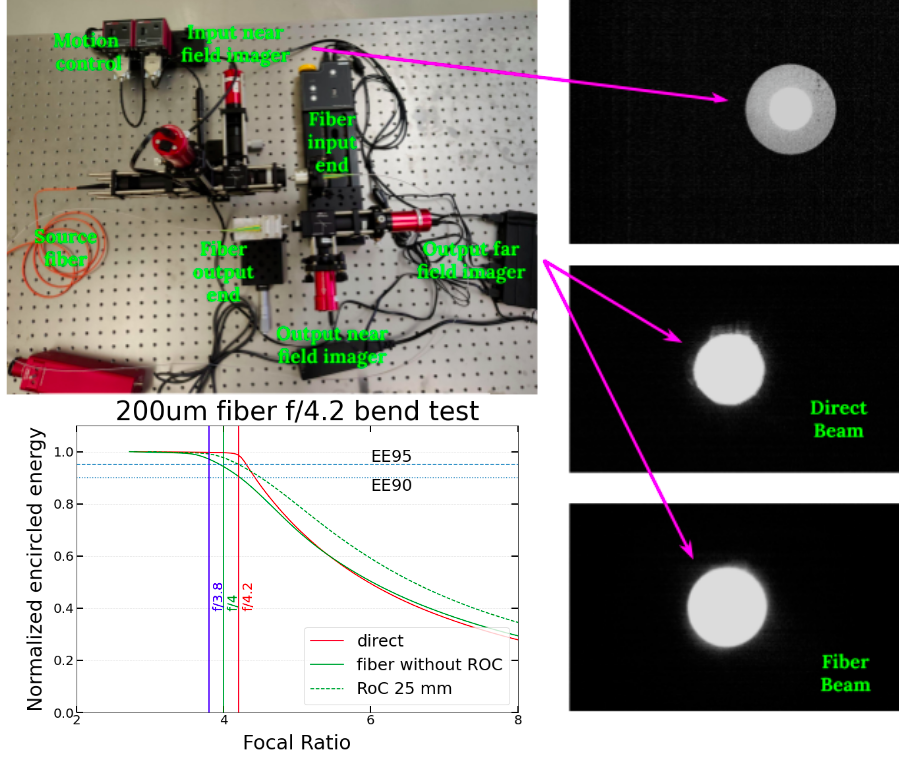}
    \caption{Focal ratio degradation performance of a sample 200 $\mu$m diameter core fibre bent full loop at 50mm diameter.}
    \label{fig:FibrePerformance}
\end{figure}

Fibre performance is measured through SAAO-Wisconsin fibre Tester (SWiFT). SWiFT is made from off-the shelf components for individual and bulk testing of focal ratio degradation and relative throughput measurement which is shown in figure \ref{fig:FibrePerformance}. It is a double-differential measurement system with two-axis motorized motion-control input module (for, e.g., mapping to the IFU and sky bundles) and one-axis motorized output module (for mapping to the pseudo-slit). The optics are 25mm or smaller. At the input of the input module, a source fibre is used to feed into optical assembly. The source fibre acts as a field-stop and is changed for different input spot diameters. A 92:8 beam-splitter splits the beam after it is collimated by a 100mm focal-length achromatic doublet. The fainter beam is reimaged onto a CMOS detector to measure the stability of the source. The brighter beam is then split again with a second 92:8 beam-splitter with the brighter beam converged by a 30mm focal-length aspheric achromat to the intermediate focus between input and output module; the fainter beam is unused. The second beam-splitter is rotated by 90 deg from the first beam-splitter to minimize polarization effects. It is used to deflect reflected light from the intermediate focus, and reimage it onto a second CMOS detector; this serves as the alignment-cum-focusing camera. A variable aperture iris following the second beam-splitter is used to tune the focal ratio at the intermediate focus.

At the injection of the output module, following the intermediate focus, the diverging beam (either from the fibre output or the direct beam) is collimated by a 14mm aspheric achromat. The beam is then split a third time, again by 92:8, with the fainter beam focused onto a third CMOS detector to image the fibre-output near-field. The pupil formed by the 14mm lens for the brighter beam is imaged onto a fourth detector which is used to measure the flux and diameter of the far-field pattern. The far-field distribution from the fibre output versus the direct beam is compared to measure the total throughput and the focal-ratio degradation.

Figure \ref{fig:FibrePerformance} shows the variation in encircled energy against focal ratio for the input and fibre output beam and associated throughput for a f/4.2 input beam over a broad (400nm - 700nm) bandpass for 200um fibre. We find the fibre output beam at EE95 is f/4 for the smallest possible bend diameter of 50mm inside the SMI cartridge. However the encircled energy profile does not follow that of the direct beam. In contrast the unbent fibre displays \textit{worse} FRD (f/3.8 at EE95) compared to fibre with the 50mm bend radius. Although it seems counter-intuitive, we believe this is due to loss of modes corresponding to wider angles. This is consistent with a diminished throughput for the bent fibre noted below.

We measured the \textit{total} fibre throughput by taking the ratio of flux within the EE99 radius of the fibre beam to the flux within the E99 radius of the direct beam. We found EE99 is at f/3.8 for the fibre output and f/4.2 for the direct beam. The total fibre throughput is 79\% and 84\% for 1.8 arcsec (400 um) and 0.9 arcsec (200 um) fibre respectively when the fibre is bent in a full loop with a 50mm diameter. Measurements of unbent fibres yield 87\% and 93\% total throughput for large and small fibres respectively. The unbent total throughput for 400 $\mu$m is lower than 200 $\mu$m core fibre which we attribute to the end termination: the glue wick back length was much smaller in the case of 200$\mu$m fibre. Since the spectrograph only accepts an f/4.2 input beam, finally we estimate the effective fibre throughput for an f/4.2 output beam. This is found to be 78\% and 71\% for small-format and large-format IFU fibres respectively. Interestingly, the relative throughput does not change with the bending condition of the fibre. In other words, the lossy modes introduced by the fibre bend are otherwise propagated (if the fibre is unbent) into a faster beam that is vignetted by the spectrograph.

\section{Summary}

We describe a new suite of DVD-cassette-size IFUs that can be used interchangeably with long-slit and multi-slit masks on SALT's Robert Stobie Spectrograph. The compact design of these SMIs ensures no modification required at the telescope focal plane assembly. Currently the first two SMIs are in fabrication and assembly phase. The fold prisms are procured and the sky-end of the large format IFU has been terminated, glued and polished. The small format IFU fibre holders for both sky and spectrograph ends are being fabricated. The process development for fibre routing and polishing the V-groove slits is complete. The modified slit-mask cassette has been tested on telescope for mechanical integrity and acceptability within the existing setup. We have developed a fibre test stand and measured performance of single 400 $\mu$m fibres to achieve 71\% throughput for an f/4.2 beam injection to the spectrograph. Higher performance of 78\% is achieved for the 200 $\mu$m fibres under similar conditions. At the time of writing the fibres for the large-format IFU are being routed and placed inside the V-groove slits, which will be followed by their termination and polishing. Following transfer of the fully fabricated large-fibre SMI into its `flight' cassette, detailed laboratory performance testing will be undertaken with SWiFT. We anticipate the large format IFU to be scheduled for commissioning in coming months. The small format IFU will follow this timeline within several months.

\acknowledgements Support for this research has been provided by the South African National Research Foundation SARChI-114555.

% References
\bibliography{report} % bibliography data in report.bib
\bibliographystyle{spiejour}

\end{document}